# Modeling SARS Spreading on Complex Networks

Huijie Yang[*], Fangcui Zhao, Zhongnan Li, Beilai Hu

*School of Physics, Nankai University, Tianjin 300071*
[*] *huijieyangn@eyou.com*

Yun Zhou

*Department of Biology, Tsinghua University, Beijing 100084*

**Abstract.** The spreading of SARS will destruct the initial network structure to a new phase, and in turn the spreading process will be weakened effectively and finally halted by this evolution of network structure. This mechanism is called immunity of contact network in this paper. What we can do is to accelerate effectively this process only.

## I. INTRODUCTION

The characteristics of SARS spreading and the effectiveness of our countermeasures attract special attentions at present time [1,2]. The contact relations between people form a complex network, called contact network in this paper. The spreading of SARS is actually a diffusion process in this complex contact network. Several relation networks have been investigated in detail in recent years, to cite examples, the mathematician's collaboration network, the actors collaborate network and the sexual contact network. [3]. The small world graphs [4] and scale-free network [5] are introduced to capture the clustering and the power-law degree distribution observed in real graphs, respectively. Various modifications of these two theoretical models have been made to consider effects such as the life of a node, the preferential attachment rule, etc. The spreading of SARS has some special features compared with information spreading. During the construction of a relation network, increasing connections between people as more as possible is a basic rule obeyed by everyone. The occurrence and spreading of SARS interrupt this rule completely. The initial structure will change and evolve to a new state due to the spreading of SARS and our countermeasures. The relating factors include isolation of a man with SARS, the control of a man with SARS-like, the infections of a man with potential SARS, etc. Hence, to obtain a reliable description of the SARS spreading we must consider two basic features carefully. One feature is the initial structure of the contact network; the other is the rule obeyed during SARS spreading.

## II. CONTACT NETWORK

In reference [6], a solution is suggested to capture the power-law degree distribution and the clustering characteristics of a real-world network simultaneously. In this paper, we propose another solution as follows,

Construct a 1-dimensional lattice by connecting sequentially $N$ nodes with periodic condition, denoted with $S_2 = \{1,2,3,\cdots N\}$. The output degree of each node is 2. Select randomly with probability $p(3)$ from the nodes with $k=2$. All the selected nodes can form a new set of nodes, denoted with $S_3 = \{n_{31}, n_{32}, \cdots\}$. Connecting sequentially these nodes with periodic condition, the output degree of each node is then 3. Iteration of this procedure, we can obtain a regular network, whose degree distribution can be adjusted by the probability $p(k)$. $k$ is the output degree.

To make the constructed network behaves scale-free we can assume that $\rho(k) = ck^{-\alpha}$. Where $\alpha \in [2,4]$, $c$ is a constant to let $\sum_{k=1}^{n} \rho(k) = 1$. $n$ is the upper limit of output degree. The ratio of nodes whose output degree is great than or equal to $m \in [2,n]$ can be written as, $f(m) = c \cdot \sum_{i=m}^{n} i^{-\alpha} = \frac{c}{1-\alpha} \cdot \left(n^{1-\alpha} - m^{1-\alpha}\right)$. The probability $p(k)$ can then be obtained as, $p(k) = \frac{f(k)}{f(k-1)} = \frac{n^{1-\alpha} - k^{1-\alpha}}{n^{1-\alpha} - (k-1)^{1-\alpha}}$.

Starting from node numbered 1 and proceeding to $N$, perform the rewiring step. For each node, consider all the connections towards other nodes and rewiring them to randomly chosen new nodes with probability $p_r$. Double connections between two nodes are forbidden. Assuming the edge from node $x$ to node $y$ is rewired to node $z$, the node $z$ should be constrained to be a special node, whose output degree is $k(z) = k(y) - 1$. $k(y)$ is the output degree of node $y$.

The output degree distribution of this constructed network obeys a power-law. At the same time, the edges are mainly localized and with probability $p_r$ an edge is rewired to a long distance node. It should capture the clustering and power-law degree distribution characteristics simultaneously.

## III. EPIDEMIC MODEL OF SARS SPREADING

(1) Select one node randomly in the constructed initial contact network as the seed of SARS (imported case). Assume contact is the way of SARS spreading. The threshold of contact times is $\mu$. If the contact time for a node with a man with SARS exceeds $\mu$ in certain duration, he will become an infected SARS.

(2) For each day, we can assign a value $\omega_{ij}$ randomly selected in $[0,1]$ to the edge between nodes $i, j$ in the contact network. $\omega_{ij}$ can act as the measure of the contact between these two nodes. For a man with SARS, with the symptom becoming more and more distinct his contact with others will decrease rapidly and finally vanish due to isolation. It can be describe with Fermi-Dirac distribution function very well, which reads, $P(t) = \frac{d}{e^{a+bt} + c}$. Where $0 \leq t \leq T_m$, $T_m$ is the duration of a SARS's staying in the contact network since his being infected. The parameters $a, b$ are sensitive to the decrease quantity and the potential time, respectively. In this model, we assign $(a,b) = (-2, 0.25)$. The parameter $d = \omega_{ij}$. Let $t = 0$, we have, $P(0) = \frac{\omega_{ij}}{e^a + c} = \omega_{ij}$, i.e., $c = 1 - e^a$.

(3) A noticeable fact is that the duration in which a certain number of contacts are accumulated is essential to the probability of being infected. The shorter this duration, the higher risk of being infected. A suitable reckon of contact measure should be, $Total(t) = \sum_{d=1}^{t} time(d) * q^{t-d}$. Where $t$ is the days, $Total(t)$ the accumulated contact measure in this duration, $time(d)$ the contact measure at the $d$'th day, and $q \in [0,1]$. If the infection power does not decrease with time, we have $q = 1$. If the infection power decreases rapidly with time, we have $q = 0$. In this model we set $q = 0.3$.

(4) As for the isolation Policy. Once a man is declared to be SARS, he will be isolated forever from the contact network, i.e., removed. The nodes that have contacted with him in the past $m$ days will be isolated instantly. This can be implemented by disconnecting their linkages with others for $m$ days. The healthy ones will then recover their linkages with others. Under this policy the structure of initial contact network may be destructed to a new phase. (5) Another factor is defiance steps. With the rapid spreading of SARS, all possible steps are implemented. One of the effects is the decrease of infection probability due to respirator, more healthful habit, improved immunity, etc., which can be described with the increase of the threshold $\mu$. The other effect is the destruction of the structure of the contact network due to the decrease of the contact measure $\omega_{ij}$, increase of isolation duration $m$, decrease of the potential life of a SARS $T_m$.

The parameters $T_m, m, \mu, \omega_{ij}$ are not constants, but dependent strongly on the accumulated number of SARS and the increase number of SARS per day, i.e., $number(t)$ and $\Delta number(t)$.

## IV. RESULTS

A contact network with 2000 nodes is constructed, the output degree distribution of which obeys $P(k) = ck^{-3} (2 \leq k \leq 13)$. The parameters $T_m, m, \mu, \omega_{ij}$ are dependent on time as follows,

$$T_m = 14 - 0.003 \times number,$$

$$m = \begin{cases} 0.3 \times number & (0.3 \times number \prec T_m) \\ T_m & (0.3 \times number \geq T_m) \end{cases},$$

$$\Delta \omega_{ij} = -0.1 \times \omega_{ij} \times \log(\Delta number),$$

$$\mu = 2.75 + 0.15 \times Log(number).$$

Fig. (1) presents the increase number per day. Duration of 45 days is simulated. In Fig. (2) the bulletins from WHO [7] about Hong Kong (21 March 2003 to 10 May 2003) is also presented to compare with our theoretical result. Due to strong statistical fluctuations, the microstructures should not contain physical information. After several peaks SARS spreading will vanish, which can be captured very well.

In our model, all the new members of SARS each day come from the people contacting with SARS. Fig. (3) presents the evolution of the infection probability, defined as the ratio of the increase number each day to the total number of nodes contact with SARS. At the beginning, the infection probability oscillates abruptly due to small number of infected nodes. Then it will decrease and vanish finally. This can be explained with the increase of the threshold $\mu$.

To describe the destruction of structure of the contact network, Fig. (4) records the evolution of the output degree distribution during the SARS spreading. The structure changes significantly, especially the number of nodes with large output degree decreases rapidly. The hubs in the networks have much higher probability of being infected and much higher probability to transfer SARS to circumstance [8]. Thus the departure of the degree distribution from the average output degree can be a measure of the defiance ability of a network itself.

In conclusion, the SARS spreading destruct the network structure, and in turn the destructed network can halt the SARS spreading effectively. This mechanics may be called immunity of contact network. What we can do is just to accelerate this evolution process.

## ACKNOWLEDGEMENTS

This work was supported by the National Science Foundation of China under Grant No. 60274051/F0303 and No. 10175036. The Post-doctor Fund of Nankai University also supported this work.

# REFERENCES

1. M. Lipstich, T. Cohen, B. Cooper, J. Robins, S. Ma, *et al.*, Science/ www.sciencexpress.org /23 May 2003 / Page 1/10.1126/science.1086616.
2. S. Riley, C. Fraser, C. A. Donnelly, A. C. Ghani, L. J. Abu-Raddad, *et al.,* Science/ www.sciencexpress.org / 23 May 2003 / Page 1/10.1126/science.1086478.
3. M. J. Newman, arXiv: cond-mat/0303516 v1, 25 Mar 2003. And references there in.
4. D.J. Watts and S.H Strogatz, Nature (London) 393,440(1998).
5. A. –L. Barabasi and R. Albert, Science 286, 509(1999).
6. J. Davidson, H. Ebel, S. Bornholdt, Phys. Rev. Lett. 88, 128701(2002).
7. WHO. http://www.who.int/csr/sars.
8. Z. Dezso, A. –L. Barabasi, arXiv: cond-mat/0107420 v2 24 Mar. 2002.

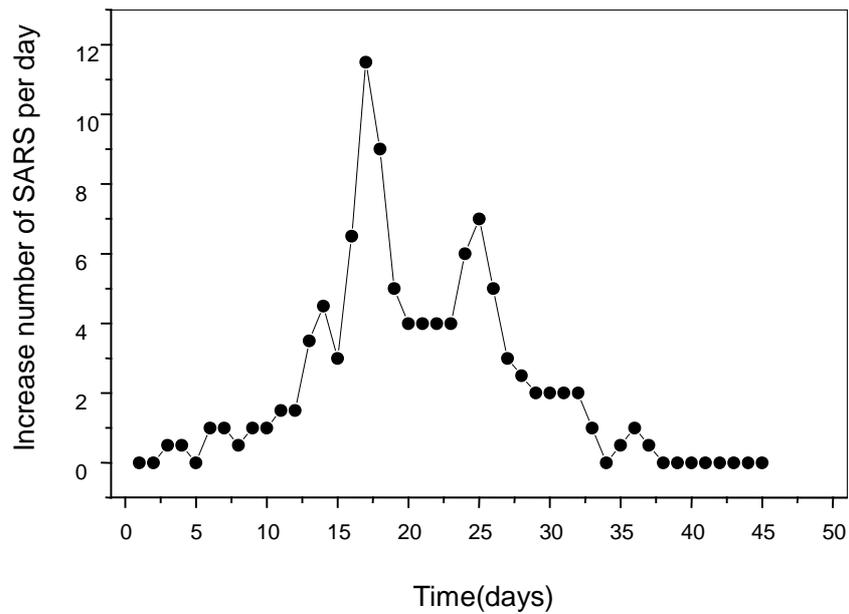

**Fig.(1)** Increase number of SARS per day. Theoretical result.

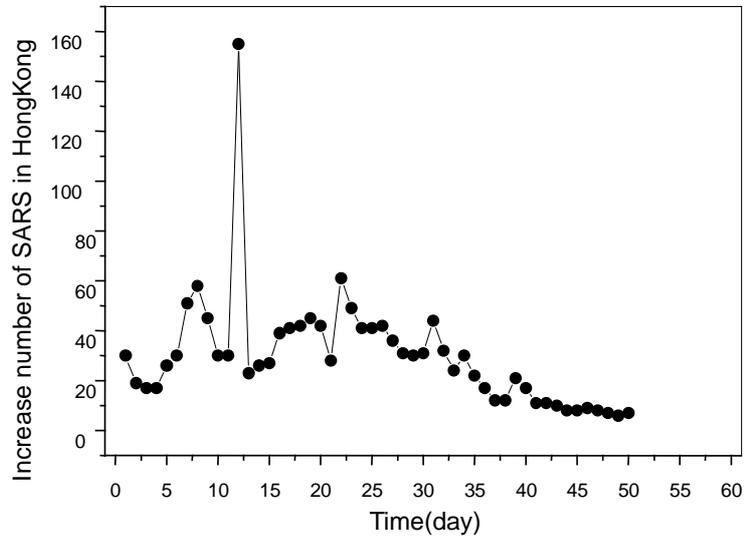

**Fig.(2)** Increase number of SARS in HongKong in duration from 1 March 2003 to 10 May 2003.

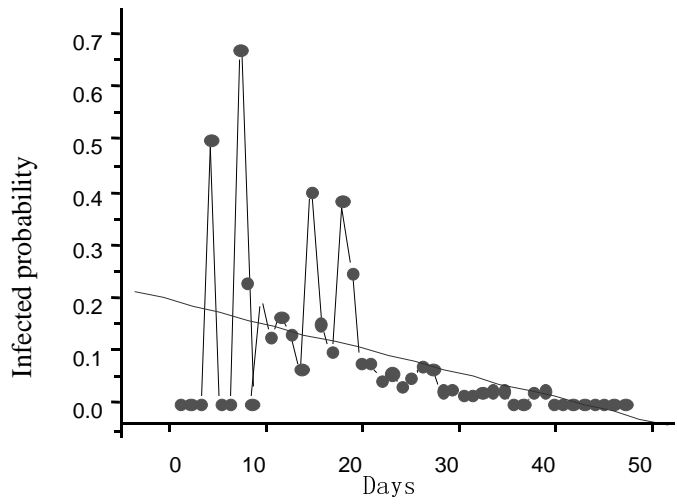

**Fig.(3)** Infected Probability. Theoretical result.

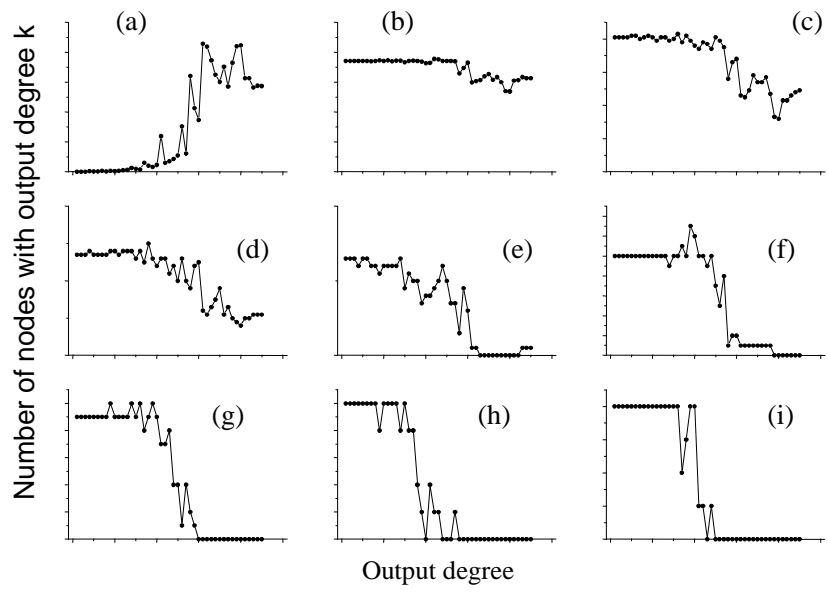

**Fig.(4)** The number of the nodes with output degree k during SARS spreading. From (a) to (i) are the results for the 0, 2,4,6,8,10,12,14,15 days, respectively.